\documentclass[onecollarge,natbib]{svjour2}
\bibpunct{[}{]}{;}{n}{}{,} 
\smartqed  
\usepackage{graphicx}
\usepackage{braket}
\graphicspath{{./figures/}}
\usepackage{amsmath}
\journalname{Archive of Applied Mechanics}
\begin{document}

\title{Structure and decay at rapid proton capture waiting points}


\author{D. Hove \and E. Garrido \and A. S. Jensen \and H. O. U. Fynbo \and D. V. Fedorov \and N. T. Zinner}


\institute{D. Hove \and A. S. Jensen \and H. O. U. Fynbo \and D. V. Fedorov \and N. T. Zinner \at
Department of Physics and Astronomy, Aarhus University, 8000 Aarhus C, Denmark \\ \email{dennish@phys.au.dk}           
\and
E. Garrido  \at
Instituto de Estructura de la Materia, IEM-CSIC, Serrano 123, E-28006 Madrid, Spain
}

\date{Received: date / Accepted: date}

\maketitle

\begin{abstract}
We investigate the region of the nuclear chart around $A \simeq 70$ from a three-body perspective, where we compute reaction rates for the radiative capture of two protons. One key quantity is here the photon dissociation cross section for the inverse process where two protons are liberated from the borromean nucleus by photon bombardment. We find a number of peaks at low photon energy in this cross section where each peak is located at the energy corresponding to population of a three-body resonance. Thus, for these energies the decay or capture processes proceed through these resonances. However, the next step in the dissociation process still has the option of following several paths, that is either sequential decay by emission of one proton at a time with an intermediate two-body resonance as stepping stone, or direct decay into the continuum of both protons simultaneously. The astrophysical reaction rate is obtained by folding of the cross section as function of energy with the occupation probability for a Maxwell-Boltzmann temperature distribution. The reaction rate is then a function of temperature, and of course depending on the underlying three-body bound state and resonance structures. We show that a very simple formula at low temperature reproduces the elaborate numerically computed reaction rate.

\end{abstract}

\section{Introduction \label{sec:intro}}

In the evolution of some exotic stellar environments the rapid proton capture (rp) process is thought to be central in determining the final nuclear distribution and abundances along with the overall time scale of the processes involved \cite{wal81,sch98}. As increasingly more protons are captured eventually the proton dripline is reached. The nuclei along the dripline are called waiting points with regards to the rp-process, and the nuclei with the longest lifetime are then called critical waiting points. The critical waiting points in the rp process above Ni are $^{64}\text{Ge}$, $^{68}\text{Se}$, and $^{72}\text{Kr}$, where $^{68}\text{Se}$ is thought to be the most important \cite{tu11}. At these critical waiting points the process is stalled and can seemingly only proceed through the much slower $\beta$-decay. However, the large pairing effect \cite{hov13} leads to an irregular shape of the proton dripline \cite{bro02}, which presents another option. If two protons were captured simultaneously, the gap in the dripline could be skipped, which would change the effective lifetime of the nuclei. 

This makes the critical waiting points ideal candidates to study from a three-body perspective. A system where the addition of one proton creates and unbound system, but the addition of two protons creates a bound system is the definition of a Borromean structure. It is very common to study such Borromean structures from a few-body perspective \cite{zhu93,hov14}, but this is usually limited to nuclei lighter than $^{38}$Ca \cite{gor95}. Our intent is to study $^{68}\text{Se} + p + p$ from a three-body perspective, and determine two- and three-body energy relations along with reaction rates for the process.

\section{Three-body formalism and structure \label{sec:3b}}

Our three-body method is the hyperspheric, adiabatic, expansion of the Faddeev equations in coordinate space. Only a brief explanation will be included here, but the interested reader can find a more detailed presentation in Ref.~\cite{nie01}. The central idea is to separate the three-body wave function into an (hyper)angular, $\Omega$, and a (hyper)radial, $\rho$, part. The angular part of the Faddeev equations are then solved as a function of $\rho$, and the angular solution is used to solve the radial part. The only thing that must be chosen to begin with is the two-body interaction between the protons and between the core and a proton. The proton-proton interaction used here is a phenomenologically established interaction presented in Eq. (5) in Ref.~\cite{gar04}. This is used for all partial waves up to an including f-waves. The core-proton interaction is a shallow Woods-Saxon potential with a central and a spin-orbit part
\begin{align}
V(r) 
=& V_C(r) +   \frac{V_0}{1 + e^{(r -R)/a}}
+ \mathbf{l} \cdot \mathbf{s} \frac{1}{r} \frac{d}{d r}  
\frac{V_0^{ls}}{1 + e^{(r -R_{ls})/a_{ls}}}, \label{eq:cenpot}
\end{align}
\noindent where the radial and diffuseness parameters are kept constant at $R=7.2$, $R_{ls}=6.3$, $a=0.65$, and $a_{ls}=0.5$ all in fm. The only free parameters are the strengths, which are used to exclude Pauli forbidden states and isolate individual, allowed orbitals at specific, desired energies. For nuclei around $A \simeq 70$ and $N \simeq Z$ recent experiments, along with shell model calculation, and comparisons with mirror nuclei indicate that the dominating orbitals are $f_{5/2}$ and $p_{3/2}$ \cite{nic14,nes14}. 

\begin{figure}
\centering
\begin{minipage}{.4\textwidth}
	\includegraphics[width=1\linewidth]{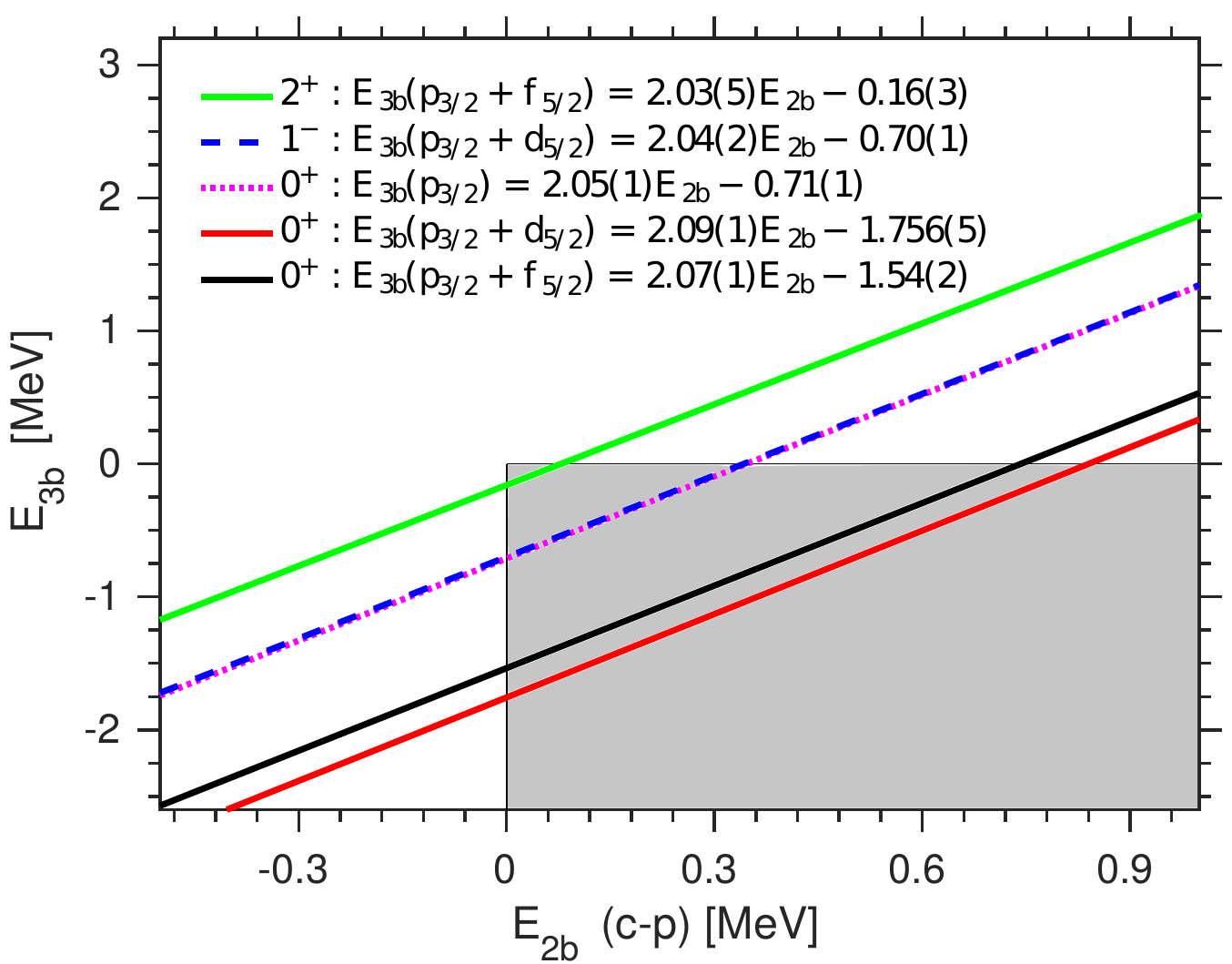}
\end{minipage}\hfill
\begin{minipage}{0.55\textwidth}
	\caption{The relationship between two- and three-body energies in $^{68}\text{Se}+p+p$. All single-particle obitals, execpt the ones indicated, have been eliminated using the two-body potentials. The grey rectangle outlines the Borromean region of interest. When two orbitals are allowed the energy has been kept identical. If the two-body energy of one orbital was increased above the other the corresponding curve would be moved up. At some point only one orbital would contribute, and the possible change in energy is therefore limited. However, negative parity states would not be limited, as both orbitals are needed to form it. \label{fig:Energy}}
\end{minipage}
\end{figure}

Our intent is to see how these orbitals affect the three-body system both in isolation and in combination. The $V^{ls}_0$ parameter is used to isolate spin-orbit partners, and $V_0$ is used to exclude the Pauli forbidden states and adjust the two-body energy of allowed states. The result is seen in Fig.~\ref{fig:Energy} for a number of orbital combinations. If two orbitals are allowed, the energy of the two are kept identical. This is an unrealistic assumption, but it does provide one limit for the energy. Taking the $0^+$ state with $p$ and $f$ waves as an example, if the energy of the $f$ orbital was increased, the black curve would move up. At some point it would coincide with the dotted curve, and it would not increase any further. This provides another limit for the energy. The same would happen for the $2^+$ state, but for the $1^-$ state, the energy would increase without limit, as both $d$ and $p$ are needed to form the negative parity state. This greatly affects the dipole transition rate, to the point where the quadrupole transition could very well be dominating, as the $d$ state most likely is very high-lying.

\begin{figure}
\centering
\begin{minipage}{.5\textwidth}
	\includegraphics[width=1\linewidth]{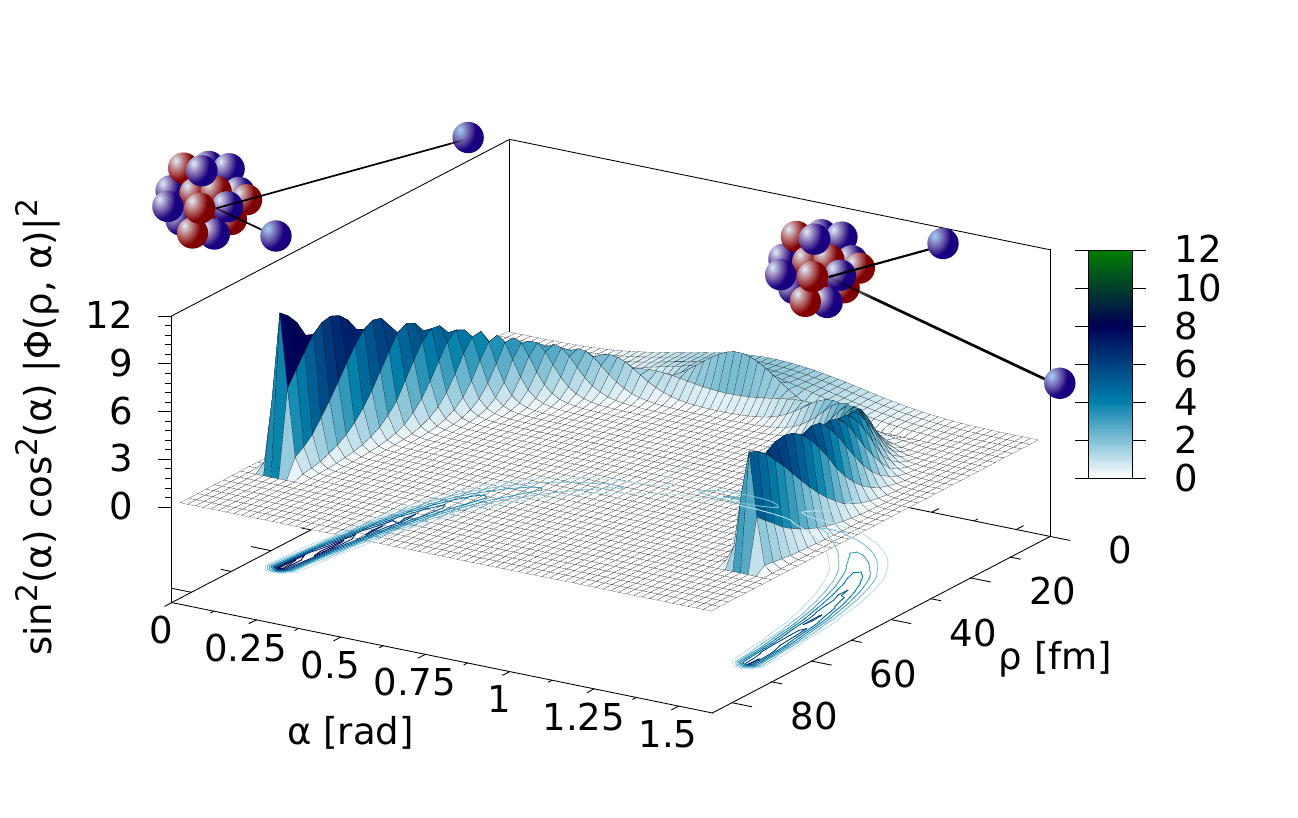}
\end{minipage}\hfill
\begin{minipage}{0.45\textwidth}
	\caption{The square of the three-body wave function of $^{68}\text{Se}+p+p$ as a function of hyperradius and hyperangle. This is for the $2^+$ state allowing both $f$ and $p$ waves with an energy of $0.64$ MeV. The illustration on top shows the structure at large $\rho$.  \label{fig:wavefunc}}
\end{minipage}
\end{figure}

Another very interesting point is the three-body structure of the system. The square of the angular wave function corresponding to the lowest angular eigenvalue is seen in Fig.~\ref{fig:wavefunc} as a function of hyperradius and hyperangle. This is for the $2^+$ state in the three-body system, allowing both $f$ and $p$ waves with an energy of $0.64$ MeV. This particular energy was chosen, as it is the recently measured proton separation energy of $^{69}$Br \cite{san14}. The structure at large $\rho$ is illustrated above each peak. One of the protons are close to the core, and the other one is far away. This particular angular wave function therefore corresponds to a sequential process, where first one and then another proton is captured. Other higher-lying angular wave functions could correspond to direct reactions or perhaps mixtures of the two. It should be noted that a specific reaction mechanism is not assumed, but all possibilities are included, as long as a sufficient number of angular eigenvalues are included. This makes the calculations rather comprehensive compared to many other few-body calculations.

\section{Reaction rate \label{sec:reac}}

In addition to the two- and three-body energy relationship, and the three-body structure and reaction mechanism we are also interested in the reaction rate, $R_{ppc}$ of three-body transition process $c + p + p \rightarrow A + \gamma$. As with the three-body formalism, detailed derivations of the formalism behind will not be included. The interested reader is referred to Ref.~\cite{hov16}. The reaction rate must here be averaged over the Maxwell-Boltzmann distribution of the temperature in the stellar environment. It can be expressed by
\begin{align}
\braket{R_{ppc}(E)} = \frac{1}{2T^3} \int E^2 R_{ppc}(E) \exp(-E/T)
 \, dE, \label{eq:ave_rate}
\end{align}
\noindent where $E$ is the three-body energy. The reaction rate $R_{ppc}$ depends most importantly on the overlap matrix element $\braket{\Psi_{0}| | \hat{\Theta}_{\ell} | | \psi_{\ell}^{(i)} }$ between the discretized three-body continuum states $\psi_{l}^{(i)}$ and the initial three-body state $\Psi_0$. This depends crucially on the electric multipole operator, $\hat{\Theta}_l$, of order $l$. This can all be calculated, but it is a rather comprehensive and inflexible calculation. 

Instead we aim to present a flexible method, which can easily be adjusted in case new information regarding the two-body energies become available. If the three-body resonances are well-separated and very narrow a Breit-Wigner shape can be assumed. In that case the integral in Eq.~(\ref{eq:ave_rate}) can be solved analytically. This results in a much simpler expression for the energy averaged reaction rate
\begin{align}
\braket{R_{ppc}(E)} =
 \frac{4 \pi^3 (2\ell+1)\hbar^5}{(\mu_{cp} \mu_{cp,p})^{3/2}}  
 \frac{\Gamma_{eff}(E_R)}{T^3} \exp(-E_R/T), \label{eq:rate_esti}
\end{align}
\noindent where the proton $\Gamma_{ppc}$ and photon $\Gamma_{\gamma}$ decay width are used to calculate the effective $\Gamma_{eff}^{-1} = (\Gamma_{\gamma}+\Gamma_{ppc})^{-1}$ decay width. In this expression the rate is dictated by the photon and proton decay widths, which depends on the short-ranged properties and the Coulomb barrier thickness respectively.

\begin{figure}
\centering
\begin{minipage}{.4\textwidth}
	\includegraphics[width=1\linewidth]{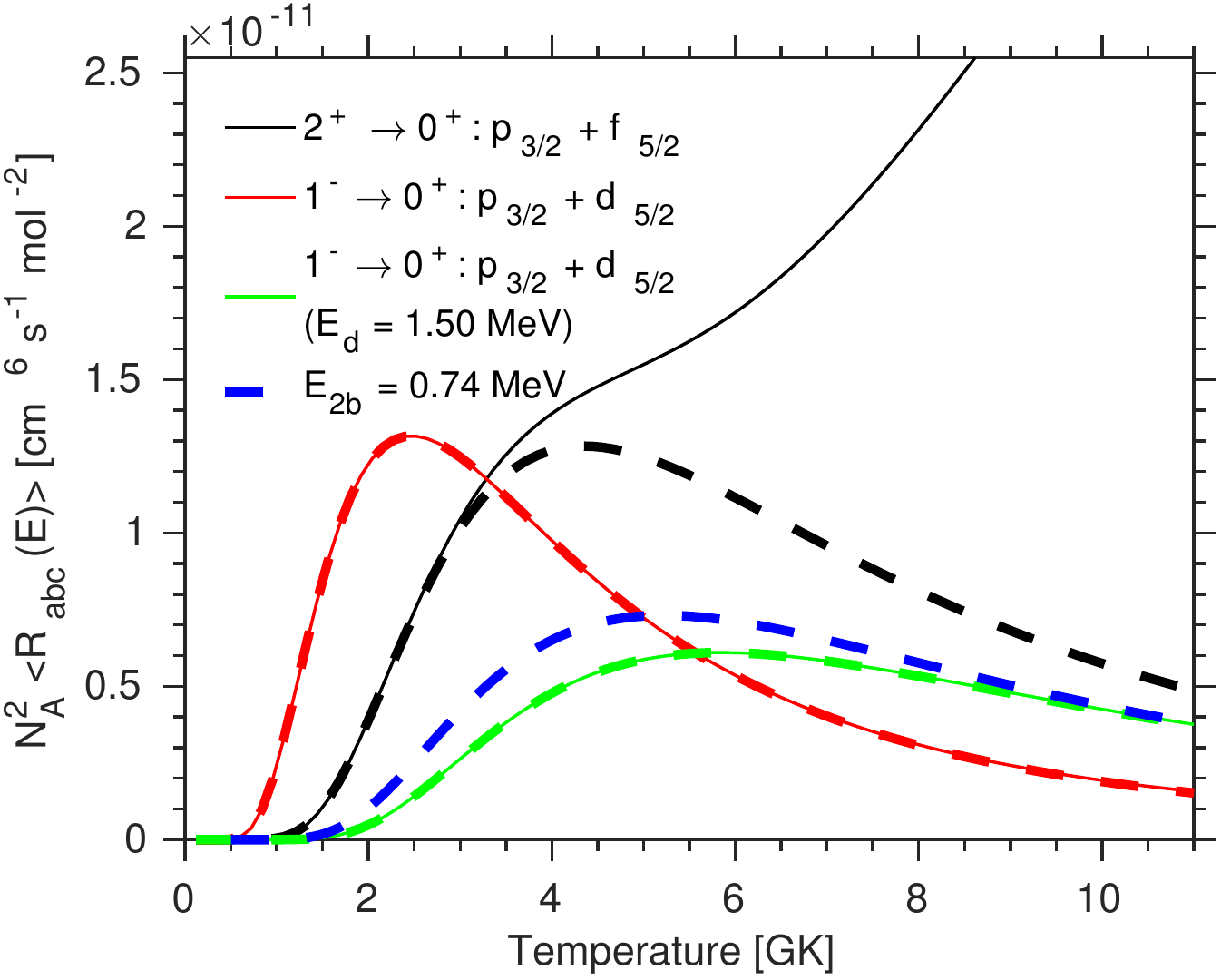}
\end{minipage}\hfill
\begin{minipage}{0.55\textwidth}
	\caption{The reaction rate for the $2^+ \rightarrow 0^+$ quadrupole transition and the $1^- \rightarrow 0^+$ dipole transition keeping the energies of the single-particle orbitals the same. The dipole transtions have been scaled down by a factor of $10^4$. The full lines are based on the full calculation of Eq.~(\ref{eq:ave_rate}), while the dashed line is based on the simple expression from Eq.~(\ref{eq:rate_esti}) using only a single three-body resonance.  \label{fig:rates}}
\end{minipage}
\end{figure}

The reaction rates for this specific energy calculated with the full rate expression are included in Fig.~\ref{fig:rates} as full lines. The result of the simplified calculation from Eq.~(\ref{eq:rate_esti}) are included as dashed lines. This is done for both the dipole (red lines) and the quadrupole (black lines) transition, where the two-body energy is 0.64 MeV for all orbitals. The dipole transition is scaled down by a factor of $10^4$. Also included is the dipole transition rate, where the energy of the $d_{5/2}$ resonance is increased to 1.5 MeV (green line). If the energy of $d_{5/2}$ is increased to 3.0 MeV, the potential becomes unable to sustain a $1^-$ resonance, and the rate decreases with many orders of magnitude, making the quadrupole transition the dominating one. Finally, the blue dashed line is the quadropole transition, calculated using Eq.~(\ref{eq:rate_esti}), where the energy of both the $p_{3/2}$ and $f_{5/2}$ resonance is increased to 0.74 MeV.

Having both single-particle orbitals at an energy of 0.64 MeV corresponds to an extreme limit, where the measured ground state of $^{69}$Br is degenerate. The other limit, where both single-particle orbitals have an energy of 0.74 MeV, is given by the Borromean region in Fig.~\ref{fig:Energy}. This energy, in combination with the simplified expression from Eq.~(\ref{eq:rate_esti}), has produced the blue dashed curve in Fig.~\ref{fig:rates}. It is thereby possible easily to predict the change in reaction rate as a response to a change in two-body energies. This is particularly useful in regions of the nuclear chart where additional experimental information is expected to be available in the near future.

The general procedure for adjusting the rates based on a change in two-body energy is as follows; first the three-body energy is determined from Fig.~\ref{fig:Energy}. Second, the shift in Coulomb potential is calculated taking into account the decay mechanism deduced from Fig.~\ref{fig:wavefunc}. This is used to calculate $\Gamma_{ppc}$ from a WKB tunneling probability. The change in $\Gamma_{\gamma}$ is negligible, as the short-range structure is unchanged under minor energy variations. Third, $\Gamma_{\gamma}$ and $\Gamma_{ppc}$ is used to calculate $\Gamma_{eff}$. Fourth and finally, Eq.~(\ref{eq:rate_esti}) is used along with $\Gamma_{eff}$ to estimate the change in reaction rate. For more details on this see Ref.~\cite{hov16}.


\end{document}